\documentclass[twocolumn,epjc3]{svjour3}
\usepackage{multirow}
\usepackage{graphicx}
\usepackage{bm}
\usepackage{dcolumn}
\usepackage{amsmath}
\usepackage{color}
\usepackage{stfloats}

\def \lleq {\lower0.9ex\hbox{ $\buildrel < \over \sim$} ~}
\def \ggeq {\lower0.9ex\hbox{ $\buildrel > \over \sim$} ~}

\def \beq  {\begin{equation}}
\def \eeq  {\end{equation}}
\def \ber  {\begin{eqnarray}}
\def \eer  {\end{eqnarray}}

\journalname{Eur. Phys. J. C}
\begin{document}
\newcommand{\newc}{\newcommand}

\newc{\be}{\begin{equation}}
\newc{\ee}{\end{equation}}
\newc{\ba}{\begin{eqnarray}}
\newc{\ea}{\end{eqnarray}}
\newc{\bea}{\begin{eqnarray*}}
\newc{\eea}{\end{eqnarray*}}
\newc{\D}{\partial}
\newc{\ie}{{\it i.e.} }
\newc{\eg}{{\it e.g.} }
\newc{\etc}{{\it etc.} }
\newc{\etal}{{\it et al.}}
\newcommand{\nn}{\nonumber}
\newc{\ra}{\rightarrow}
\newc{\lra}{\leftrightarrow}
\newc{\lsim}{\buildrel{<}\over{\sim}}
\newc{\gsim}{\buildrel{>}\over{\sim}}
\title{Constraints of Dark Energy at High Redshift}
\author{Qiping Su\thanksref{e1,addr1}
        \and
        Rong-Gen Cai\thanksref{addr2} %etc.
}
\thankstext{e1}{e-mail: sqp@hznu.edu.cn}
\institute{Department of Physics, Hangzhou Normal University, Hangzhou, 310036, China \label{addr1}
           \and
           State Key Laboratory of  Theoretical Physics, Institute of
Theoretical Physics, Chinese Academy of Sciences, P.O. Box 2735,
Beijing 100190, China \label{addr2}
}

\date{Received: date / Accepted: date}
\maketitle
\begin{abstract}
Constrains of dark energy (DE) at high redshift from current and mock future observational data are obtained.
It is found that present data give poor constraints of DE even beyond redshift z=0.4, and mock future 2298 type Ia supernove data only give a little improvement of the constraints.
We analyze in detail why constraints of DE decrease rapidly with the increasing of redshift.
Then we try to improve the constraints of DE at high redshift.
It is shown that the most efficient way is to improve the error of observations.
\end{abstract}

\section{Introduction}
The current expansion of the universe is found to be in accelerating
\cite{Riess:1998cb,Perlmutter:1998np} and it is believed that the
dark energy (DE), whose equation of state $w_{de}=p_{de}/\rho_{de}$
is less than $-1/3$, plays the role to drive the accelerated
expansion of the universe. Plenties of DE models have been proposed
\cite{Peebles:2002gy,Copeland:2006wr,Steinhardt:1999nw,Capozziello:2003tk,Cai:2007us,Cai:2009ht,Cai:2010uf}
but the origin of DE is still unknown. We have only known a little
about DE from observations.

Several parametrization methods of
$w_{de}$ have been proposed to fit with observations to get
constraints of $w_{de}$. Usually in parametrization models $w_{de}$ is
assumed to be a simple function of redshift z, e.g., the CPL
parametrization~\cite{Chevallier:2000qy,Linder:2002et}:
$w_{de}(z)=w_0+w_az/(1+z)$,  and the parametrization of redshift
expansion~\cite{Cooray:1999da,Huterer:2000mj}: $w_{de}(z)=w_0+w_zz$.
It has been found that $w_{de}$ is very close to -1 and
is varying very slowly (if it is dynamical). But these fitting
results depend on the parametrization forms of $w_{de}(z)$ used.
Moreover, one parametrization form of $w_{de}(z)$
could only approximate the real $w_{de}$ well in a very limited
region of redshift.
So fitting results from a single model should not be used to analyzing behaviors of DE at both low and high redshift.
There are also ``model-independent" methods\cite{Huterer:2002hy,Huterer:2004ch,Wang:2009sn,Holsclaw:2010nb},
in which the redshift of data is usually divided into several bins and in each redshift bin $w_{de}$ is assumed to be a simple function of z.
All ``model-independent" methods show that constraints of $w_{de}$ at higher redshift are much weaker than that at low redshift~\cite{Sullivan:2007pd,Zhao:2007ew,Serra:2009yp,Cai:2010qp,Hojjati:2009ab,Su:2012ci}.
It is important and necessary to get good constraints of $w_{de}$ at high redshift to reveal the nature of DE,
such as the dynamical behaviors of DE.

In this paper, we'd like to analyze constraints of $w_{de}$ at high redshift.
Since present constraints of DE are from observational data, we will mainly analyze effects of observational data on constraining DE at high redshift.
We will start with type Ia supernovae (SnIa) data.
Almost all present data for DE are related to the comoving distance and the most data of this distance type are from SnIa.
Commonly used DE data other than the distance type are CMB data and Hubble parameter data only.
And main information in CMB data can be extracted to distance priors (such as the shift parameter R date):
there is only a little difference between constraints of DE from full CMB data and from the distance priors\cite{Wang:2013mha}.
The distance priors are related to distances, i.e., one can convert the CMB data to data of distance type.
So our analyses will be mainly based on the distance type of data, and study why distance type of data give poor constraints of DE at high redshift.

The paper is organized as follows.
In section II, constraints of $w_{de}$ at low redshift and high redshift from present and mock future data are obtained. In section III, we analyze the properties of $w_{de}$ at high redshift in detail and try to find out reasons of the poor constraints.
In section IV, we try to improve constraints of $w_{de}$ at high redshift by adding number of date points, decreasing error of observational data and adding other type of data (i.e., Hubble parameter data).

\section{Constraints from distance type of data}

At present, almost all observational data for DE are the distance type, such as data of SnIa, BAO parameter A and CMB shift parameter R, which are related to a comoving distance: $r_c=\int dt/a=\int dz/H(z)$.
At first, we'd like to estimate the effect of SnIa data (since it gives the most date points for DE) on constraints of DE, especially constraints at high redshift. Since a single SnIa date set has poor constraints on DE, the BAO parameter A will also be adopted to alleviate the degeneracy between equation of state ($w_{de}$) and dimensionless density energy ($\Omega_{de}$) of DE.

\begin{figure}[t]
  \includegraphics[width=3.5in,height=2.3in]{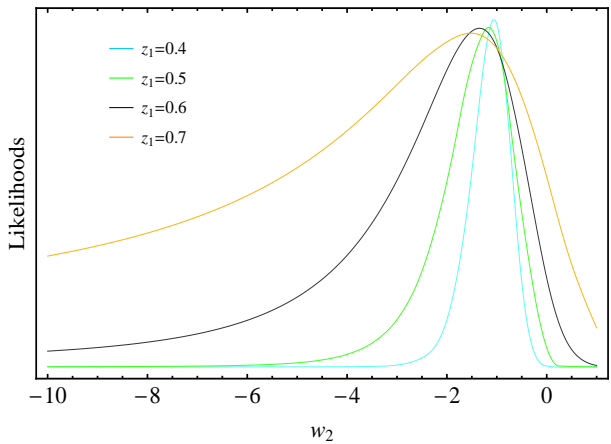}
  \caption{The likelihoods of $w_2$ from present data, with divided position $z_1$=0.4, 0.5, 0.6, 0.7, respectively.} \label{fig1}
\end{figure}

We will use the UBE method to get constraints of $w_{de}$ from present and future mock data.
We divide the redshift of data into two bins and set
\be
w_{de}(z)=\left\{
\begin{array}{cc}
w_1~,& 0\leq z\leq z_1 \\
w_2~,&z_1<z
\end{array}\right.,\label{w}
\ee
where $w_1$ and $w_2$ are just constants and $z_1$ is the divided position of low and high redshift.
In the numerical calculations we will set the prior $w_2>-20$, or $w_2$ will run to large minus value in MCMC procedure and the lower error of $w_2$ will be extremely large. This will be discussed in detail in the next section.
For each date set the figure of merit (FoM)~\cite{Wang:2008zh,Su:2011ic} is also calculated, which is defined as:
\be
{\rm FoM}=\left[{\rm det}C(w_1,w_2)\right]^{-1/2}~,\label{fom}
\ee
where $C(w_1,w_2)$ is the covariance matrix of $w_1$ and $w_2$ after
marginalizing out all other parameters.
In general, FoM is used to estimate the the goodness of the data in constraining $w_{de}$.

Our calculations show that the correlation between $w_1$ and $w_2$ is very small,
i.e., the errors of $w_1$ and $w_2$ obtained can be treated as independent with each other.
So errors of $w_2$ ($w_1$) just represents constraints of $w_{de}$ at high (low) redshift.

\begin{figure}[t]
  \includegraphics[width=3.5in,height=2.3in]{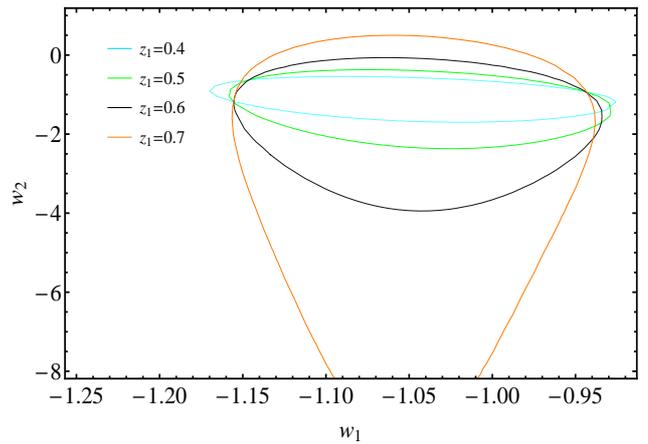}
  \caption{68\% C.L. contour plot in $w_1\sim w_2$ plane from present data, with divided position $z_1$=0.4, 0.5, 0.6, 0.7, respectively.} \label{fig2}
\end{figure}

\begin{table}[b!]
\begin{centering}{
\begin{tabular}{cccc}
\hline
$z_1$&$w_1$&$w_2$&FoM\\
\hline
0.4&  $-1.04_{-0.09-0.17}^{+0.03+0.14}$ & $-1.04_{-0.43-1.05}^{+0.36+0.67}$ &28.47 \\
\hline
0.5&  $-1.04_{-0.08-0.17}^{+0.07+0.14}$ & $-1.19_{-0.92-2.43}^{+0.68+1.15}$ &12.68 \\
\hline
0.6&  $-1.04_{-0.08-0.18}^{+0.06+0.13}$ & $-1.33_{-2.87-18.52}^{+1.30+2.13}$ &3.52 \\
\hline
0.7&  $-1.04_{-0.09-0.17}^{+0.06+0.12}$ & $-1.50_{-14.84-18.47}^{+2.14+2.44}$ &2.36 \\
\hline
\end{tabular}}
\caption{The best-fitted values and their 68.3\% and 95.4\% C.L. errors of $w_1$ and $w_2$ from present observational data, the divided positions are $z_1=0.4, 0.5, 0.6 , 0.7$, respectively. } \label{present}
\end{centering}
\end{table}

\subsection{Constraints from present date set}

Here we get constraints of $w_{de}$ from Union2.1 SnIa data \cite{Suzuki:2011hu} and BAO Parameter $A$ date from \cite{Eisenstein:2005su}.
To estimate constraints of $w_{de}$ at different high redshift bins, the divided position $z_1$ will be set as 0.4, 0.5, 0.6 and 0.7, respectively.
The best-fitted parameters and their 68.3\%, 95.4\% confidence level (C.L.) errors are obtained by using the Markov chain Monte Carlo (MCMC) method, which are shown in Table~\ref{present}.
We have following conclusions:

1. $w_{de}$ at high redshift ($z>0.4$) are highly unconstrained by present data,
especially compared with $w_{de}$ at low redshift.
With $z_1=0.4$, the average 95.4\% C.L. error for $w_2$ (noted as $2\bar{\sigma}(w_2)$ ) is $$2\bar{\sigma}(w_2)=(0.67+1.05)/2=0.86$$ and $2\bar{\sigma}(w_1)$ is only about 0.15, as shown in Table~\ref{present}.

2. Errors of $w_{de}$ at high redshift (i.e., errors of $w_2$) increase rapidly with the divided point $z_1$, especially the lower errors of $w_{2}$.
As shown in Fig.~\ref{fig1},
$z_1$ is larger, the likelihood of $w_{2}$ decreases more slowly with the decreasing of $w_2$.
We will analyze this phenomenon in detail in the next section.

3. Effect of the divided position $z_1$ on errors of $w_{de}$ at low redshift (i.e., $w_1$) is very weak.
As shown in Table.~\ref{present}, with the increasing of $z_1$ variations of errors of $w_1$ are negligible.
It implies that present date set at low redshift ($z<0.4$) is sufficient and constraints on $w_{de}$ in this region is strong.

4. Here, values of FoM can be used to estimate goodness of constraints of $w_{de}$ at high redshift.
FoM from Eq.(\ref{fom}) is proportional to the inverse area of the $1\sigma$ error ellipse in the $w_1\sim w_2$ plane.
As shown from the contour plots in Fig.~\ref{fig2} and
errors of $w_1$ in Table~\ref{present}, the decreasing of FoM (with respect to the increasing of $z_1$) is due to the increasing of errors of $w_2$.
So value of FoM is larger, errors of $w_2$ are smaller, and vice versa.
Indeed, FoM increases rapidly with the decreasing of $z_1$, as shown in Table~\ref{present} and Fig.~\ref{R}.

\subsection{Constraints from future data}

%%%%%%%%%%%%%%%%%%%%%%%%%%%%%%%%%%%%%%%%%%%%%%%%%%%%%%%%%%%%%%%%%%%%%%%%%%%%%%%%%%%%%%%%%%%
To estimate the constraints of $w_{de}$ at high redshift from future data,
we fit the 2-binned UBE model with 2298 simulated SnIa data (denoted as elementary date set in this paper),
which contain 1998 SnIa data with redshift $0.1<z<1.7$ from a SNAP-like JDEM survey
and 300 SnIa data with $z<0.1$ from the NSNF \cite{Kim:2003mq,Ealet:2002vj}.
To alleviate the degeneracy between $\Omega_{m0}$ and $w_{de}$\cite{Maor:2000jy},
the date of BAO Distance Parameter $A$~\cite{Eisenstein:2005su} will also be included.
To simulate the mock SnIa data, we assume the fiducial model as $w_{de}(z)=-1$. The error of
the distance modulus  ($\delta$) for each supernova is set as $0.13$~\cite{Holsclaw:2010nb}.
To estimate the effect of the divided position $z_1$, we have set $z_1$ as 0.4, 0.6, 0.8 and 1.0, respectively.

%%%%%%%%%%%%%%%%%%%%%%%%%%%%%%%%%%%%%%%%%%%%%%%%%%%%%%%%%%%%%%%%%%%%%%%%%%%%%%%%%%%%%%%%%%%
\begin{table}[b!]
\begin{centering}
{\begin{tabular}{cccc}
\hline
$z_1$&$w_1$&$w_2$&FoM\\
\hline
0.4&  $-0.96_{-0.04-0.09}^{+0.03+0.07}$ & $-1.12_{-0.24-0.55}^{+0.17+0.30}$ &119.65 \\
\hline
0.6&  $-0.97_{-0.06-0.12}^{+0.03+0.08}$ & $-1.14_{-0.42-1.31}^{+0.27+0.45}$ &33.23 \\
\hline
0.8&  $-0.96_{-0.07-0.15}^{+0.04+0.08}$ & $-1.11_{-0.69-4.03}^{+0.43+0.73}$ &11.68 \\
\hline
1.0&  $-0.95_{-0.08-0.15}^{+0.02+0.06}$ & $-1.11_{-1.88-18.73}^{+0.89+1.28}$ &5.89 \\
\hline
\end{tabular}}
\caption{The best-fitted values and their 68.3\% and 95.4\% C.L. errors of $w_1$ and $w_2$ from mock future data, the divided positions are $z_1=0.4, 0.6, 0.8 , 1.0$, respectively.} \label{future}
\end{centering}
\end{table}

As shown in Table~\ref{future},
with much more date points, the errors of $w_{de}$ from future data are smaller than that from the present data.
E.g., here $2\bar\sigma(w_2)=0.43$ for $z>0.4$, which is just a half of that from the present data (i.e., 0.86).
Errors of $w_{de}$ at high redshift (i.e., $w_2$) also increase rapidly with $z_1$, e.g.,
here $2\bar\sigma(w_2)$ for $z>0.6$ is 0.88.
It also shows that constraints of $w_{de}$ beyond $z=1$ are extremely weak.
On the other hand, constraints of $w_{de}$ at low redshift (i.e., constraints of $w_1$) are also improved,
but not as much as that of $w_2$.

In all, present data gives poor constraints of $w_{de}$ at high redshift(e.g., $z>0.4$),
and the mock future 2298 SnIa data do not give sufficient improvement on constraints of $w_{de}$ at high redshift.
\section{Analyzing constraints}
Now we try to answer two questions:
\subsection{Why data of distance type give poor constraints on $w_{de}$ at high redshift?}
Reasons are:

1. Date points in a redshift bin will constrain $w_{de}$ in this bin and in lower redshift bins,
but can not constrain $w_{de}$ in higher redshift bins, e.g., $w_{de}$ in the lowest redshift bin can be
constrained by all date points.
Moreover, the redshift is larger the corresponding distance is farther, and it will be harder to measure the distance.
And the farther a supernova is, the harder it can be detected. While the most date points for DE are from SnIa.
In all, there are much less efficient date points in higher redshift bins.

2. In higher redshift, the effect of dark energy on the luminosity distance \be D_l(z)=(1+z)\int_0^z dz'/H(z')\label{DL} \ee is much less. Here $H^2(z')\sim \rho_{de}(z')+\rho_{dm}(z')$.
With respect to the increase of redshift z, energy density of DE ($\rho_{de}$) evolves very slowly, while energy density of DM ($\rho_{dm}$) increases very rapidly since $\rho_{dm}\sim(1+z)^3$.
In higher redshift bin, $\rho_{de}$ is much less than $\rho_{dm}$.
So constraints of DE at high redshift from distance type of data will be weaker.

\subsection{Why lower errors of $w_{de}$ at high redshift bins are extremely large}
This can be shown by simple calculations as
follows. In a flat FRW universe the Fridemann equation reads: \be
H^2(x)=\frac{1}{3}[\rho_{m0}(1+x)^3+\rho_{de}(x)],\label{H} \ee
where $
\rho_{de}(x)=\rho_{de0}e^{3\int_0^x\frac{1+w_{de}(y)}{1+y}dy}.\nonumber $

 \begin{table*}[!t]
\be \delta D_l(z)=\left\{
\begin{array}{cc}
0~,& 0\leq z\leq z_{i} \\
 -\frac{3}{2}(1+z)\int_{z_i}^z\frac{\delta w_i}{H(x)}\Omega_{de}\ln{(\frac{1+x}{1+z_i})}dx~,& z_i<z\leq z_{i+1}\\
 -\frac{3}{2}(1+z)
 \left[\int_{z_i}^{z_{i+1}}\frac{\delta w_i}{H(x)}\Omega_{de}\ln{(\frac{1+x}{1+z_i})}dx+\int_{z_{i+1}}^{z}\frac{\delta w_i}{H(x)}\Omega_{de}\ln{(\frac{1+z_{i+1}}{1+z_i})}dx\right]~,&z_{i+1}<z
 \end{array}\right. \label{dd}
 \ee
 \end{table*}

The redshift is divided into n bins and in each bin $w_{de}$ is set as a constant.
Supposing there is a tiny variation of $w_{de}$ in the
$i^{th}$ bin ($\delta w_i$) and no variations of $w_{de}$ in all other bins, from
Eq.(\ref{DL}) and (\ref{H})  one gets
\be \delta H^2(x)=\left\{
\begin{array}{cc}
 0~,&0\leq x\leq z_i\\
\rho_{de}\ln{(\frac{1+x}{1+z_i})}\delta w_i~,& z_i< x\leq z_{i+1} \\
\rho_{de}\ln{(\frac{1+z_{i+1}}{1+z_i})}\delta w_i~,& z_{i+1}<x
 \end{array} \right.
 \ee
and Eq.(\ref{dd}) for $\delta D_l(z)$, where $\Omega_{de}=\rho_{de}/(\rho_{de}+\rho_{dm})$.

One can see that $\delta w_i$ always appears together with
$\Omega_{de}$ in Eq.(\ref{dd}).
At higher redshift (i.e., larger i), with the same $\delta w_i$ $\Omega_{de}$ is much smaller and $\delta D_l(z)$ is also smaller.
It indicates that the redshift
is higher, to lead to the same variation of $D_l$ the variation of $w_i$ must be larger, i.e., $D_l(z)$ is less sensitive to $w_i$ at higher redshift.
In all,
constraints on $w_{de}$ from the luminosity distance are weaker in
higher redshift bins.

As shown in Eq.(\ref{dd}), $\delta D_l$ in the $i^{th}$ bin is related to an
integration, which is mainly determined by $\Omega_{de}$ and the width of this redshift bin.
The width of a redshift bin is larger, the integration will be larger and so will be $\delta D_l$.
In higher redshift bins $\Omega_{de}$ is smaller since
\be
\Omega_{de}(z_i<z<z_{i+1})\sim (1+z)^{3w_i}~,
\ee
so $\delta D_l$ will be smaller. Note that $\Omega_{de}$ is also dependent on $w_i$

\begin{table}[b!]
\begin{centering}
{\begin{tabular}{cccc}
\hline
multiple&$w_1$&$w_2$&FoM\\
\hline
$\times2$ & $-0.96_{-0.08-0.13}^{+0.03+0.06}$ & $-1.33_{-1.95-18.64}^{+0.82+1.13}$ &6.64 \\
\hline
$\times3$ & $-1.00_{-0.05-0.11}^{+0.02+0.06}$ & $-1.01_{-0.86-3.48}^{+0.54+0.93}$ &15.41 \\
\hline
$\times4$ & $-1.00_{-0.04-0.08}^{+0.03+0.06}$ & $-1.20_{-0.72-2.49}^{+0.49+0.89}$ &22.03 \\
\hline
$\times5$ & $-0.99_{-0.04-0.06}^{+0.02+0.05}$ & $-1.00_{-0.47-1.27}^{+0.37+0.65}$ &92.48 \\
\hline
\end{tabular}}
\caption{The best-fitted values and their 68.3\% and 95.4\% C.L. errors of $w_1$ and $w_2$ from simulated SnIa data.
The numbers of SnIa data are $2, 3, 4$, and $5$ times of the $2298$ elementary data, respectively. The divided position $z_1$ is set to $1.0$.}\label{number}
\end{centering}
\end{table}

Now let's answer the question.
In a high redshift bin if $w_i$
is much smaller than $-1$, $\Omega_{de}$ in this bin will be so
small that $\delta D_l$ keeps small even the bin is
wide and  $\delta w_i$ is large, i.e. with different small values of $w_i$ one
gets almost the same $D_l$. In this case, the likelihood
of $w_i$ will be very flat for $w_i\ll-1$ as shown
in Fig.~\ref{fig1}. Thus the lower errors of $w_i$ at high
redshift bins are always extremely large. In general, the constraint of
$w_{de}$ in the last redshift bin is the weakest.

\section{Improving constraints}

According to previous results and analyses,
we try to improve the constraints of $w_{de}$ at high redshift.
Three methods will be implemented and their efficient will be compared.

\subsection{Adding the number of SnIa data}

In the section 2.2 it is shown that $\sim2300$ mock future SnIa data still give poor constraints of $w_{de}$
beyond $z\sim0.6$. Here we try to use more mock SnIa data, with 2, 3, 4 and 5 times of the 2298 elementary data.
The proportional distribution of redshift in these multiple date sets is the same as the elementary date set.
The divided position $z_1$ of the two bins is now set to 1.0~.
The results are shown in Table~\ref{number}.
It shows that FoM increases with the increasing of number of date points, but the increase is not very efficient.
With 5 times number of the elementary SnIa date set (about 11500 supernovae), constraints of $w_{de}$ beyond $z=1$ are still weak: $2\bar\sigma(w_2)=0.96$.

\subsection{Improving accuracy of the data}

%%%%%%%%%%%%%%%%%%%%%%%%%%%%%%%%%%%%%%%%%%%%%%%%%%%%%%%%%%%%%%%%%%%%%%%%%%%%%%%%%%%%%%%%%%%
\begin{table}[b!]
\begin{centering}
{\begin{tabular}{cccc}
\hline
$\delta$&$w_1$&$w_2$&FoM\\
\hline
0.13 &  $-1.00_{-0.05-0.11}^{+0.02+0.06}$ & $-1.01_{-0.86-3.48}^{+0.54+0.93}$ &15.41 \\
\hline
0.10&  $-1.00_{-0.04-0.07}^{+0.01+0.05}$ & $-1.01_{-0.55-1.52}^{+0.41+0.72}$&30.57\\
\hline
0.05&  $-1.00_{-0.01-0.04}^{+0.01+0.03}$ & $-1.01_{-0.24-0.55}^{+0.22+0.39}$ &386.48 \\
\hline
0.02&  $-1.00_{-0.00-0.01}^{+0.00+0.00}$ & $-1.00_{-0.09-0.19}^{+0.08+0.16}$&2478.16 \\
\hline
\end{tabular}}
\caption{The best-fitted values and their 68.3\% and 95.4\% C.L. errors of $w_1$ and $w_2$ with errors in distance modulus $\delta=0.13, 0.10, 0.05, 0.02$, respectively.
The number of SnIa used is $3$ times of 2298 elementary SnIa data.
The divided positions  $z_1$ is set to $1.0$.}\label{error}
\end{centering}
\end{table}

Increasing the number of SnIa data seems not efficient enough in improving constraints of $w_{de}$ at high redshift.
It is expected that with the increase of number of data the statistical errors of the data will be decreased.
Moreover, the systematic errors of the data will be improved by future observations~\cite{Weinberg:2012es}.
To estimate effects of the error $\delta$ of distance modulus of supernovae, here we set $\delta=0.13, 0.1, 0.05, 0.02$, respectively.
The number of the SnIa data used here will be $3$ times of the elementary data and $z_1=1.0$ is set.

The results in Table~\ref{error} show that FoM increases very rapidly with the decrease of $\delta$.
With $\delta=0.02$, $2\bar\sigma(w_2)=0.18$ for $z>1.0$, which is a good constraint of $w_{de}$ at high redshift.

Of course, the errors of other distance type of observations are also improved continually.
For example, the $1\sigma$ error of CMB shift parameter R from WMAP9 is 0.016\cite{Hinshaw:2012aka}, which decreased to 0.009 from Planck\cite{Ade:2013zuv}.
This parameter is defined as:
$$R=\sqrt{\Omega_{m0}}\int_0^{z_s}\frac{dz}{E(z)},~~~~~  E=H/H_0$$
which is related to the distance of decoupling epoch ($z_s\sim 1091$ is the redshift of decoupling).

\begin{figure}[t]
  \includegraphics[width=3.5in,height=2.2in]{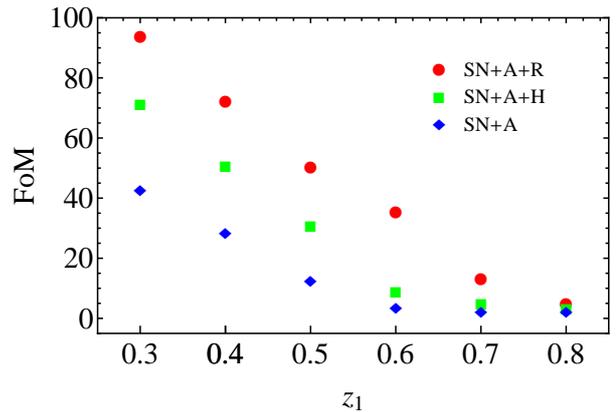}
  \caption{Values of FoM from SN+A+R (dots), SN+A+H (squares) and SN+A (diamonds), where SN is for Union2.1 SnIa date, A is for the BAO date, R is for the shift parameter date and H is for 22 Hubble parameter data. The abscissa is for the divided position $z_1$.} \label{fig3}
\end{figure}

Now We combine shift parameter R from WMAP9 with Union2.1 and BAO data to get constraints of $w_{de}$.
Main results are shown in Table~\ref{R}. In Fig.~\ref{fig3}, FoM for this date set are plotted to compare with other date sets.
With the addition of shift parameter R, $2\bar\sigma(w_2)=0.28$ for $z_1=0.4$, without this parameter $2\bar\sigma(w_2)=0.86$ for $z_1=0.4$.

%%%%%%%%%%%%%%%%%%%%%%%%%%%%%%%%%%%%%%%%%%%%%%%%%%%%%%%%%%%%%%%%%%%%%%%%%%%%%%%%%%%%%%%%%%%
\begin{table}[t!]
\begin{centering}
{\begin{tabular}{cccc}
\hline
$z_1$&$w_1$&$w_2$&FoM\\
\hline
0.4&  $-1.03_{-0.07-0.16}^{+0.08+0.15}$ & $-0.86_{-0.25-0.54}^{+0.20+0.37}$ &72.22 \\
\hline
0.5&  $-1.03_{-0.05-0.14}^{+0.08+0.15}$ & $-0.82_{-0.35-0.84}^{+0.25+0.45}$ &50.45 \\
\hline
0.6&  $-1.03_{-0.05-0.14}^{+0.07+0.14}$ & $-0.74_{-0.50-1.39}^{+0.30+0.48}$ &35.56 \\
\hline
0.7&  $-1.04_{-0.03-0.11}^{+0.10+0.15}$ & $-0.67_{-0.76-2.78}^{+0.34+0.53}$ &13.13 \\
\hline
\end{tabular}}
\caption{The best-fitted values and their 68.3\% and 95.4\% C.L. errors of $w_1$ and $w_2$ from data of SN+A+R, the divided positions are $z_1=0.4, 0.5, 0.6 , 0.7$, respectively.}\label{R}
\end{centering}
\end{table}

The addition of a single date point R gives a big improvement of FoM and constraints of $w_{de}$ at high redshift,
main reasons are:\\
1) The error of R date (0.016) is very small, compared with the present errors of SnIa. As shown in previous results, the error of date is smaller, the constraints of $w_{de}$ will be better. Moreover, the corresponding redshift for the parameter R is $z\sim1091$, it gives good constraints of $w_{de}$ in all redshift bins.\\
2) To involve the date of parameter R, the second redshift bin should be ($z_1$,1091).
As analyzed in section 3.2, the bigger a redshift bin is, the effect of $w_{de}$ on $D_l$ will be bigger.
So the constraint of $w_2$ from parameter R is good.\\
3) The shift parameter R date can further alleviate the degeneracy between $w_{de}$ and $\Omega_{de}$.

\subsection{Combined with other type of data}

There is few type of date other than distance type, the typical one is the Hubble parameter data which constrain $w_{de}$ directly: there is no integration between the Hubble parameter and $w_{de}$.

Here we combine 22 Hubble parameter data\cite{Farooq:2013dra} with the Union2.1 and BAO data to show effects of the Hubble parameter data.
The results are shown in Table \ref{h} and Fig.~\ref{fig3}.
With the addition of the 22 Hubble date points, FoM is improved, but not as much as that with shift parameter R.
Though one needs not to have integration in fitting Hubble data,
the errors of Hubble data is much bigger than that of shift parameter R date and the width of second bin related to the Hubble data is much smaller than that of R date.

\begin{table}[h!]
\begin{centering}
{\begin{tabular}{cccc}
\hline
$z_1$&$w_1$&$w_2$&FoM\\
\hline
0.4&  $-1.01_{-0.07-0.14}^{+0.06+0.11}$ & $-1.01_{-0.32-0.77}^{+0.23+0.40}$ &50.93 \\
\hline
0.5&  $-1.01_{-0.06-0.13}^{+0.06+0.12}$ & $-1.05_{-0.57-1.40}^{+0.32+0.51}$ &30.89 \\
\hline
0.6&  $-1.02_{-0.07-0.13}^{+0.06+0.11}$ & $-1.10_{-0.94-4.67}^{+0.46+0.71}$ &9.12 \\
\hline
0.7&  $-1.02_{-0.08-0.15}^{+0.05+0.10}$ & $-1.10_{-1.70-16.84}^{+0.61+0.84}$ &4.96 \\
\hline
\end{tabular}}
\caption{The best-fitted values and their 68.3\% and 95.4\% C.L. errors of $w_1$ and $w_2$ from data of SN+A+H, the divided positions are $z_1=0.4, 0.5, 0.6 , 0.7$, respectively.}\label{h}
\end{centering}
\end{table}

\section{Summary}

We have analyzed constraints of $w_{de}$ at high redshift from current and future observations.
It was shown that at higher redshift, constraints of $w_{de}$ from all observational date sets are much weaker.
The present data give poor constraints on $w_{de}$ beyond $z\sim0.4$, whose average 95.4\% C.L. error is 0.86.
With the future 2298 mock data, average 95.4\% C.L. error of $w_{de}$ beyond $z\sim0.4$ is about 0.43.
We have carefully analyzed why constraints of DE at high redshift from observational data are so poor.
Since almost all DE data is of distance type, our analyses are mainly based on distance type of data.
The analyses show that it is hard to get good constraints of $w_{de}$ at high redshift from distance type of data.
Then we tried to improve constraints of DE at high redshift by adding more numbers of future mock SnIa data, improving the errors of observational data, and combining other type of data (Hubble parameter data) with distance types of data.
It was shown that improving the error of observations is the most efficient way.
About $6900$ SnIa data with observational errors $\sigma=0.02$ can constrain $w_{de}$ beyond $z=1$ within 0.1 at 68.3\% C.L. and within 0.2 at 95.4\% C.L.~.

At present, many projects for DE are in progress or in planning.
It is expected that a great deal of date points will be released.
The distance types of date will still be the most important ones for revealing the nature of DE.
To reveal the nature of DE, good constraints of $w_{de}$ at high redshift is required.
Our results show that much more date points will be needed, but improvements for accuracy of the observations are much more efficient and necessary.
While the most present observational data for DE are distance type or can be converted into distance type,
another way to improve the constraints of $w_{de}$ is developing more other types of data.

{\bf Acknowledgements:}

This work was supported in part by the
National Natural Science Foundation of China (No. 11247008, No. 11175225 and No. 11147186), in part by Zhejiang Provincial Natural Science Foundation of China under Grant No. LQ12A05004, and in part by the Ministry of
Science and Technology of China under Grant No. 2010CB833004 and No. 2010CB832805.

\end{document}